\documentclass[twocolumn,showpacs,preprintnumbers,amsmath,amssymb]{revtex4}

\usepackage{graphicx}

\begin{document}
\title{Slow Mass Transport and Statistical Evolution of An Atomic Gas Across the \\ Superfluid-Mott Insulator Transition}
\author{Chen-Lung Hung}
\author{Xibo Zhang}
\author{Nathan Gemelke}
\author{Cheng Chin}
\address{The James Franck Institute and Department of Physics, \\ The University of Chicago, Chicago, IL 60637, USA}

\date{\today}

\begin{abstract}
We study transport dynamics of ultracold cesium atoms in a two-dimensional optical lattice across the superfluid-Mott insulator transition based on
\emph{in situ} imaging. Inducing the phase transition with a lattice ramping routine expected to be locally adiabatic, we observe a global mass redistribution which requires a very long time to equilibrate, more than 100 times longer than the microscopic time scales for on-site interaction and tunneling. When the sample enters the Mott insulator regime, mass transport significantly slows down. By employing fast recombination pulses to analyze the occupancy distribution, we observe similarly slow-evolving dynamics, and a lower effective temperature at the center of the sample.
\end{abstract}


\maketitle

The thorough understanding of atomic interactions in optical lattices
provides a testing ground to investigate hypothetical models widely
discussed in condensed matter and many-body physics
\cite{Bloch_review,Lewenstein07}. Because of the simplicity and tunability of the underlying Hamiltonian, research on optical lattices generates new
fronts to perform precise, quantitative comparison between
theoretical calculations and measurements. This new class of
``precision many-body physics'' has generated tremendous interest
in recent years to locate the superfluid (SF) to Mott insulator (MI) phase boundaries \cite{Bloch2002, spielman08, Trotzky2009}, described by the Bose-Hubbard model \cite{fisher, Jaksch98}, and to characterize Mott and band
insulators in Fermi gases \cite{Tilman, Bloch}, described by the
Fermi-Hubbard model \cite{Hubbard}. Many new, exotic quantum phases
in optical lattices have also been proposed \cite{Lewenstein07}, even in the absence of counterparts in condensed matter physics.

As promising as the precise characterization of quantum phases is, fundamental assumptions such as the thermal dynamic equilibrium of the sample should be investigated. Since the preparation of quantum gases generally involves ramping up the lattice potential, dynamics are an inseparable part of all optical lattice experiments. Very slow equilibration processes have been reported in one-dimensional optical lattices \cite{Kinoshita2006} and have been suggested by the observation of long-lived repulsively bound pairs \cite{Winkler2006} and doublons \cite{Strohmaier09} in three-dimensional lattices. Prospects of non-equilibrium dynamics in optical lattices have also attracted much interest recently. Mass and entropy transport in the optical lattices can provide a wealth of information to characterize the underlying quantum phases \cite{Altman05,Mun07}. Dynamic passage across a phase transition can lead to the proliferation of topological defects in the optical lattices \cite{Cucchietti07}.

\begin{figure}[t]
\includegraphics[width=3.4 in]{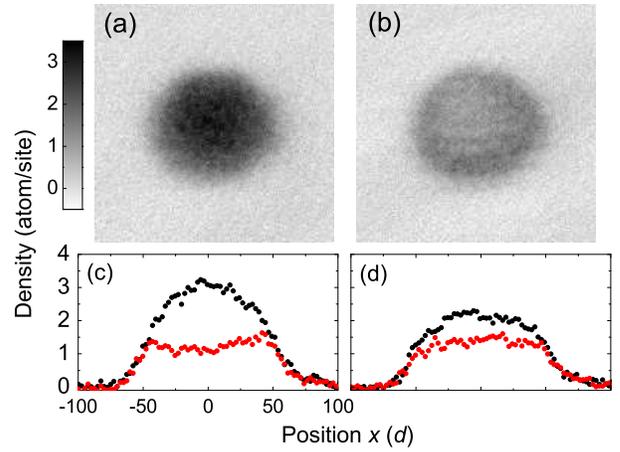}
\caption{(Color online) Averaged absorption images and density cross sections
of $N=2 \times 10^4$ cesium atoms in a monolayer of 2D optical lattice.
After ramping the lattice in 150~ms to a lattice depth
$V_f=13~E_R$, (a) shows the sample immediately after the ramp. In
(b), an additional fast recombination pulse removes atoms in sites
of occupancy three or more. (c) shows the average density cross sections of (a) (black circles) and (b) (red circles). (d) shows the average density cross sections of the samples with additional 800~ms hold time after the ramp, without (black circles) and with the recombination loss pulse (red circles). Image size is $(106~\mu$m$)^2$ = (200~sites)${^2}$.} \label{fig1}
\end{figure}

In this letter, we study global dynamics of ultracold atomic
gases in a monolayer of two-dimensional (2D) optical
lattice. After ramping up the lattice potential, we observe both mass transport and statistical distribution of atomic occupancy in the lattice. Mass transport is directly seen from \emph{in situ} density profiles, while occupancy statistics is probed by inducing loss in sites of three or more atoms using a fast three-body recombination (see Fig.~\ref{fig1}). Both processes show intriguing behavior at times much longer than microscopic time scales for atomic interaction and tunneling. 

We begin the experiment with a $^{133}$Cs quantum gas in a 2D
optical trap. Details on the preparation of the quantum gas and
optical lattice loading procedure can be found in Ref.~\cite{Hung08}
and Ref.~\cite{Gemelke09}, respectively. In brief, a nearly pure
Bose condensate is loaded into a 2D optical dipole trap, formed by two
orthogonally crossed beams on the $x-y$ plane and a one-dimensional
vertical optical lattice of 4~$\mu$m spacing which confines the whole sample in a single ``pancake''-like lattice site \cite{Gemelke09}. Using microwave
tomography, we find $\sim 95\%$ of the atoms are loaded into a singe
pancake trap. The remaining $\sim 5\%$ in the neighboring sites do
not contribute to the main results reported in this letter. The trap
vibration frequencies are $(\omega_x,\omega_y,\omega_z)=2\pi\times(11,13,1970)$~Hz, and the
cloud temperature is $T=11$~nK. The ratios
$\hbar\omega_i/k_BT=(0.05, 0.06, 9)$ indicate the sample is
two-dimensional. After 2D trap loading, we adjust the atomic
scattering length $a$ by ramping the magnetic field to a designated
value, typically, $B=20.7$~G where $a=200$~$a_B$ and $a_B$ is the
Bohr radius. At this field, the three-body recombination loss rate
is at the Efimov minimum \cite{Kraemer06}.

We introduce a 2D optical lattice by slowly turning on
retro-reflections of the crossed dipole beams which add a square
lattice potential with lattice spacing $d=532$~nm and a weak
contribution to the envelope confinement characterized by a mean radial frequency $\sqrt{\omega_x\omega_y}=2\pi(1+V/82E_R)\times 12$~Hz, where
$E_R=k_B\times 64$~nK is the recoil energy and $V$ is the lattice
depth. Care is taken to equalize lattice depths in the $x$ and
$y$ directions by balancing the lattice vibration frequencies to within $5\%$.
Based on the lattice vibration frequency measurement, we calculate tunneling
$t$ and on-site interaction $U$ numerically from the band structure
in a homogeneous 2D lattice.

We ramp on the lattice depth following $V(\tau)=V_f(1+\gamma)/[1+\gamma
e^{4(\tau-\tau_c)^2/\tau_c^2}]$ \cite{Clark04}, preceded by a 30~ms
linear ramp from $0$ to 0.4~$E_R$. The final depth $V_f$ is reached
at time $\tau=\tau_c$ and $\gamma$ is chosen such that
$V(0)=0.4$~$E_R$. After the ramp, the sample is held in the lattice
for a hold time $\tau_{hold}$. The adiabaticity parameter of the
ramp is given by $\alpha=\hbar|\dot{t}|/t^2$; slow
ramps with $\alpha < 1$ suggests that local
equilibrium of the system is preserved \cite{Gericke07, Clark04}.

We obtain the \emph{in situ} density profile of the sample by absorption imaging normal to the $x-y$ plane. After a hold time $\tau_{hold}$, we first switch the magnetic field to $B=17.7$~G ($a = 40 a_B$) and then turn off the 2D lattice 100~$\mu$s before the imaging, reducing the on-site peak density by a factor of 30, in order to mitigate any density dependent loss during the imaging process \cite{loss}. The atomic density is measured with a spatial resolution of 1.3~$\mu$m using a long working distance ($34$~mm) commercial microscope objective. The strength and duration of the imaging pulse are chosen to keep the travel distance of the atoms due to the radiation pressure from the imaging beam small compared to the depth of focus, while maintaining a good signal-to-noise ratio.


\begin{figure}[t]
\includegraphics[width=3.4 in]{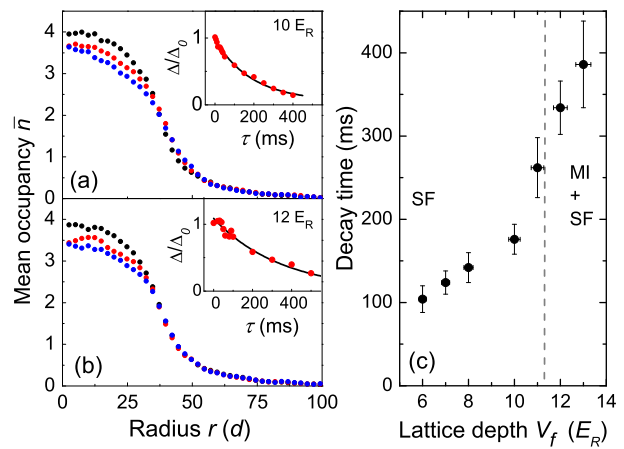}
\caption{(Color online) Evolution of the density profile after a short lattice ramp. Following a $\tau_c=20$~ms ramp to $V_{f}=10$~$E_R$ ($U/t=11$, $\alpha<0.6$), (a) shows the radial density profiles measured after hold times of 0~ms (black circles), 200~ms (red circles) and 500~ms (blue circles). Inset shows the time evolution of $\Delta$, normalized to the initial value $\Delta_0=\Delta(0)$ (solid circles), see Eq.~(1), and the single exponential fit. (b) shows the profiles measured at $V_{f}=12$~$E_R$ ($U/t=20$, $\alpha < 1$). The fitted decay times at different depths of $V_f$ are shown in (c), where the dashed line marks the critical lattice depth, see text.} \label{fig2}
\end{figure}

Our first step to study the global dynamics is to watch how the
density profile comes to equilibrium after a lattice ramp. Here we employ
a ramp which is locally adiabatic, but is fast enough to induce
detectable mass flow. An example is shown in Fig.~\ref{fig2} (a), where after a $\tau_c=20$~ms ramp to $V_f=10~E_R$ ($U/t=11$, $\alpha <0.6$) the sample of $N=2 \times 10^4$ atoms at scattering length $a=200~a_B$
gently expands and the peak density slowly decreases. This deformation
is consistent with the increase of repulsive atomic interaction in
stronger lattice confinement.

To quantify the rate of mass redistribution, we define the root-mean-square deviation of a density profile at hold time $\tau$ from equilibrium as
\begin{equation}
\Delta(\tau)=\left\{ \sum_i[\bar{n}_i(\tau)-\bar{n}_{eq,i}]^2\right\}^{1/2}, \\
\label{N}
\end{equation}
\noindent where the sum goes over lattice sites enclosing the sample. $\bar{n}_i(\tau)$ is the mean occupancy of site $i$ at hold time $\tau$, obtained by averaging over an annular area centered on the cloud, containing site $i$, and with width $1.3~\mu$m \cite{Gemelke09}. $\bar{n}_{eq,i}$ is the mean occupancy of site $i$ at equilibrium, which we obtain from samples that cease to evolve after long hold time of $\tau_{hold}= 500 \sim 800$~ms.

At lattice depths $V_f < 10~E_R$, the sample shows
a weak breathing mode oscillation in the first 50~ms of hold time. After 50~ms, $\Delta(\tau)$ can be fit by single exponential decays with
time constants $> 100$~ms. When the lattice depth reaches $V_f = 11$~$E_R$ or higher, the mass flow slows down significantly, see Fig.~2 (b) and (c), suggesting that the mass transport is suppressed in this regime. The crossover behavior near $V_f = 11$~$E_R$, where $U/t \approx 15$, is consistent with a recent observation of the suppression of superfluidity at $U/t=16$ in a 2D optical lattice \cite{spielman08}, and quantum Monte Carlo calculations,
predicting that the SF-MI transition at the tip of the $n=1$ Mott lobe occurs at $U/t \approx 16.74$ in 2D \cite{Capo08}. For $V_f > 13~E_R$, even
slower dynamics require much longer hold time and the slow loss from
three-body recombination limits our ability to determine the mass
redistribution time scale \cite{3bodyrate}.

The slow dynamics throughout the SF-MI regime indicate that the
global thermalization is much slower than the microscopic time
scales. Indeed, in the range of $V_f=6\sim 13~E_R$, tunneling to
neighboring sites occurs in $\tau_t=\hbar/zt = 0.6\sim3$~ms, where
$z=4$ is the coordination number of the 2D square lattice.

In the second experiment, we investigate the evolution of occupancy statistics. For this, we develop a scheme to determine the fraction of sites with three or more atoms by inducing a fast three-body recombination loss, and comparing the density profiles with and without the loss. For cesium atoms, extremely fast three-body loss can be induced by jumping the magnetic field near an Efimov resonance \cite{Kraemer06}, where the loss happens much faster than atoms tunnel.

We induce the recombination loss at $V_f=13$~$E_R$ by jumping the magnetic field to $B=2$~G for a duration of 1~ms before imaging at $B=17.7$~G. The $1/e$ time of the field switching is below $100~\mu$s. During the switching, the magnetic field from the eddy currents is measured by microwave spectroscopy and compensated by a controlled overshoot of currents in the magnetic coils. At 2~G, the three-body loss rate is as high as $(20~\mu$s$)^{-1}$ for 3 atoms in one site, much faster than the tunneling rate $1/\tau_t=(3$~ms$)^{-1}$, and the 1~ms pulse is sufficient to remove all the atoms that could participate in the recombination process.

We analyze the dynamics of on-site statistics by first ramping the
lattice in $\tau_c=$300~ms to $V_f=$13~$E_R$ ($U/t=41$, $\alpha <
0.1$) at scattering length $a=$310~$a_B$ and then holding the sample for
up to 800~ms. Here, the lattice ramp is slow enough to ensure negligible subsequent mass flow. Density profiles at different hold times, with
and without the recombination pulse, are shown in Fig.~3 (a-c). A
larger fractional loss occurs at the central part of the sample
where the density is higher, as expected; there is no apparent loss
in the wing. When the sample is held longer in the lattice, we
observe a smaller fractional loss, which suggests that fewer sites are
found with three or more atoms. 

\begin{figure}[t]
\includegraphics[width=3.2 in]{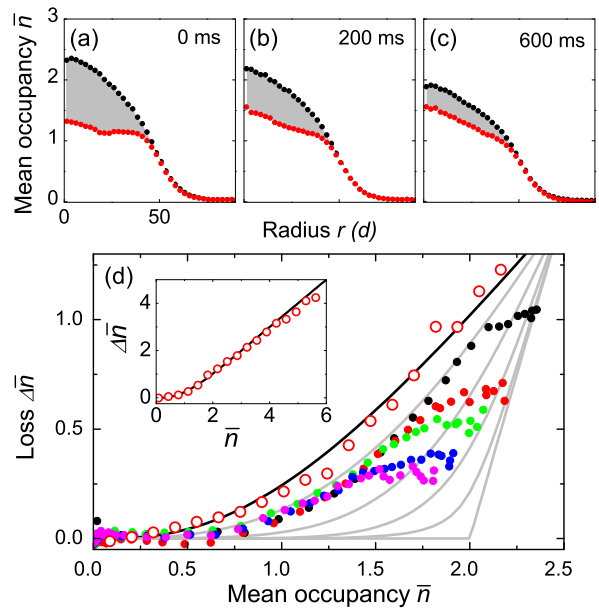}
\caption{(Color online) Evolution of the on-site statistics in the
Mott insulator. ($N=1.6\times10^4$, $\tau_c = 300$~ms,
$V_f=13$~$E_R$, $U/t=41$ and $\alpha < 0.1$). Upper figures show the
density profiles of the samples held in the final depth
$V_f=13$~$E_R$ for $\tau_{hold}=$ (a) 0~ms, (b) 200~ms, (c) 600~ms and then imaged with (red circles) and without (black circles) the recombination
pulse. Shaded areas mark the loss fractions. (d) shows the loss $\Delta\bar{n}$ versus mean occupancy $\bar{n}$ measured after different hold times (solid circles): 0~ms (black), 200~ms (red), 400~ms (green), 600~ms (blue) and 800~ms (magenta). Gray lines are the loss derived from an insulator model, see text, assuming
$k_BT/U$=1 (higher curve), 0.5, 0.3, 0.2 and 0 (lower curve). The black
line, derived from the Poisson distribution, is in good agreement
with an ideal gas measurement (open circles). The inset shows an
extended view.} \label{fig3}
\end{figure}

The evolution of the statistics is best shown in Fig.~3 (d), where
the atom loss, $\Delta \bar{n}$, is induced by the recombination
pulse after different hold times. A dramatic difference is seen near
the center with mean occupancy near $\bar{n}=2$. Here the loss
fraction reaches $\Delta \bar{n}/\bar{n}=50\%$ immediately after the
ramp, and it slowly declines to merely $15\%$ after a hold time of $\tau_{hold}=$800~ms.

To quantitatively model the loss, we assume, starting with $n$ atoms in one
site, ($n$~modulo~3) atoms remain after the pulse. To test this
model, we prepare an ideal 2D gas by tuning the magnetic field to
$B=17.1$~G, where $a\approx 0~a_B$. We then quickly ramp on the
lattice to $30~E_R$ in 10~ms to freeze the on-site occupancy and
perform the measurement with the loss pulse. For non-interacting
particles, we expect the occupancy obeys Poisson distribution. The
calculated atom loss, see black solid lines in Fig.~3 (d) and the
inset, is in good agreement with our measurement.

Recombination losses measured with interacting samples and slow lattice ramps, on the other hand, deviate from the Poisson model toward lower values for all mean occupancies (see Fig.~3 (d)). This is a general characteristic of the strongly interacting gas.

To gain further insight into the occupancy statistics in an insulator, we compare our measurement with an analytic model based on a grand canonical ensemble \cite{Ho07,Gerbier07}. In deep lattices with $t \ll U,k_B T$, the probability for occupancy $n$ can be written as $P_n=Q^{-1} {e^{-\beta (H_n-\mu n)}}$, where $H_n\approx (U/2) n(n-1)$, $\beta=1/k_BT$, $\mu$ is the local chemical potential and $Q=\sum_n e^{-\beta (H_n-\mu n)}$ is the
grand partition function. The mean occupancy is then $\bar{n}=\sum
nP_n$ and the loss is modeled as $\Delta \bar{n}=\bar{n}-\sum P_n
(n$~mod~$3)$. Calculations for $k_BT/U=1, 0.5, 0.3, 0.2$ and $0$ are
plotted in Fig.~3(d). For $\bar{n}<2.5$, all curves show smaller loss than does the Poisson model. An insulator at lower temperature experiences fewer losses because thermal fluctuation is reduced. At zero temperature, loss only occurs when the occupancy $n\geq3$ is unavoidable, or equivalently, $\bar{n}>2$.

Surprisingly, our loss measurements do not follow the model with a
uniform temperature for up to 800~ms hold time. Describing the
deviation from a constant temperature contour by an effective local
temperature $T_{eff}(r)$, we find the center of the cloud has a
lower $T_{eff} \sim 6$~nK, while for the wing $\sim$20~nK even after
$800$~ms of hold time. This persistent temperature variation across
the sample suggests that the heat flow is insufficient to establish
a global thermal equilibrium even after 800~ms hold time. This may
be aggravated by a large heat capacity of the atoms in the wing.

Both the slow mass and heat flows observed in this work raise the
issue of describing quantum gases in optical lattices using a
thermodynamic model. We suspect that the slow dynamics is partially
due to our large sample size of (100 sites)$^2$ and the
dimensionality of our system, and partially associated with the
critical behavior of the system. Across the SF-MI transition, the
sample enters the quantum critical regime, where long equilibration
times are expected \cite{Cucchietti07, Zakrzewski09}. Other
interesting mechanisms include the long life time of the exited
doublon \cite{Strohmaier09}, which could slow down statistical
redistribution of occupancies while supporting mass transport.
Moreover, the slow recombination loss during the lattice ramp and
hold time preferentially removes atoms at the center of the sample,
which could lead to the radial temperature gradient observed in our
system.

In summary, we show that the \emph{in situ} density profiles of
atoms in a 2D optical lattice provide a viable tool for investigating
dynamic processes induced by chemical potential and temperature
imbalance. In both cases, we find equilibration times much longer than
the microscopic tunneling time scale. Further investigation into these
processes and the relevance of our observation to the quantum
dynamics in the critical regime will be reported in the future.

We thank T.L. Ho, J. Freericks, D.-W. Wang, Q. Zhou and N. Trivedi for helpful
discussions and are grateful to P. Scherpelz for carefully reading the manuscript. This work is supported by NSF Award No. PHY-0747907 and under ARO Grant No. W911NF0710576 with funds from the DARPA OLE Program, and the Packard foundation. N.G. acknowledges support from
the Grainger Foundation.


\begin{thebibliography}{99}
\bibitem{loss} This procedure suppresses systematic distortions of the
density measurement. In previous works \cite{Gemelke09,ther09}, our
imaging was performed at low magnetic fields, where recombination
and radiative losses preferentially reduce the density at the center
and can enhance the plateau feature in deep lattices.
\bibitem{3bodyrate} At $V_f = 13~E_R$ and scattering length $a=200$~$a_0$, three body loss rate for 3 atoms per site is $\sim 0.4~s^{-1}$.
\end{thebibliography}
\end{document}